\documentclass[twocolumn,prl,preprintnumbers]{revtex4}

\usepackage{dcolumn}
\usepackage{bm}
\usepackage{amsmath}
\usepackage{graphicx}
\usepackage{mathrsfs}
\usepackage{amssymb}
\usepackage{amsfonts}
\usepackage{hyperref}
\usepackage{color,soul}

\begin{document}

\setlength{\textfloatsep}{2pt}
\setlength{\intextsep}{2pt}
\title{Exploring the quantum critical behaviour in a driven Tavis-Cummings circuit}
\author{M. Feng$^{1}$}
\email{mangfeng@wipm.ac.cn}
\author{Y. P. Zhong$^{2}$}
\author{T. Liu$^{1,3}$}
\author{L. L. Yan$^{1}$}
\author{W. L. Yang$^{1}$}
\author{J. Twamley$^{4}$}
\email{jtwamley@ics.mq.edu.au}
\author{H. Wang$^{2,5}$}
\email{hhwang@zju.edu.cn}
\affiliation{$^{1}$ State Key Laboratory of Magnetic Resonance and
Atomic and Molecular Physics, Wuhan Institute of Physics and
Mathematics, Chinese Academy of
Sciences, Wuhan, 430071, China \\
$^{2}$ Department of Physics, Zhejiang University, Hangzhou, 310027, China \\
$^{3}$The School of Science, Southwest University of Science and
Technology, Mianyang 621010, China \\
$^{4}$ARC Centre for Engineered Quantum Systems, Department of Physics
and Astronomy, Macquarie University, NSW 2109, Australia\\
$^{5}$Synergetic Innovation Center of Quantum Information and Quantum Physics, University of Science and Technology of China, Hefei, Anhui 230026, China}

\begin{abstract}

Quantum phase transitions play an important role in many-body systems and have been a research focus in
conventional condensed matter physics over the past few decades. Artificial atoms, such as superconducting qubits
that can be individually manipulated, provide a new paradigm of realising and exploring quantum phase transitions by engineering an on-chip
quantum simulator. Here we demonstrate experimentally the quantum critical behaviour in a highly-controllable
superconducting circuit, consisting of four qubits coupled to a common resonator mode.
By off-resonantly driving the system to renormalise the critical spin-field coupling strength,
we have observed a four-qubit non-equilibrium quantum phase transition in a dynamical manner,
i.e., we sweep the critical coupling strength over time and monitor the four-qubit scaled moments
for a signature of a structural change of the system's eigenstates.
Our observation of the non-equilibrium quantum phase transition, which is in
good agreement with the driven Tavis-Cummings theory under decoherence,
offers new experimental approaches towards exploring quantum phase transition related science, such as scaling behaviours,
parity breaking and long-range quantum correlations.

\end{abstract}
\maketitle

In a quantum phase transition (QPT) \cite{Sachdev:1413518,Zhang2013,greentree,hartmann,anglakis,mebrahtu,zhangx,zhangj},
the quantum system displays non-analytic behaviour which is reflected by
a discontinuous change in a property of the ground state or the structure of the excited
states, when a system parameter traverses a critical point.
In many cases this discontinuous change has a cusp-like character
surrounding which quantum fluctuations dominate and novel phenomena can be explored.
Quantum phase transitions are studied in a variety of naturally-grown
condensed matter materials such as conductors, superconductors and magnets.
With the introduction of well-controlled quantum elements,
ranging from cold atoms, photons and trapped ions to Josephson-junction qubits,
it becomes possible to engineer a quantum simulator, an ordered arrangement of the above-mentioned quantum elements,
to mimic and investigate the properties of complex interacting quantum materials.
Achieving a QPT using fine-tuning knobs available in an experimentally
accessible Hamiltonian presents the first step towards engineering such a simulator for exploring QPT-related
physics in few or many-body interacting quantum systems.

Recently, there has been extensive interest to investigate a QPT in the Dicke model \cite{Dicke:1954bl} using
artificially engineered systems both experimentally \cite{Baumann:2010js} and theoretically \cite {Nagy:2010dr,Bastidas:2012fr}.
As another paradigm to investigate light-matter interactions, the Tavis-Cummings (TC)
model \cite {TC} is an integrable variant of the Dicke model, which also yields significant interests
covering a wide range of configurations such as the multi-mode resonator \cite{Retzker:2007ew}
and the TC-lattice \cite{Knap:2010is}. The TC model is derived from the Dicke model
in the rotating-wave approximation which is valid when the spin-field coupling is weak
in comparison to other characteristic frequencies of the system.
It is generally understood that a QPT can occur in the Dicke model,
rather than in the TC model, with the former critical spin-field coupling required equal to
the geometric mean of the spin and field resonance frequencies. However,
most laboratory-achievable spin-field couplings can only
reach the strengths that are many orders smaller than the Dicke critical coupling strength.
Even for some systems with ultrastrong couplings \cite {nie2010, forn}, the Dicke coupling strength is still unreachable.
As a result, achieving the Dicke QPT with current laboratory techniques requires
additional assistance. For example, it was shown that an external drive in a
cold atom system leads to the Dicke Hamiltonian in the rotating frame, yielding a non-equilibrium QPT \cite{Baumann:2010js}.

Since most artificially engineered quantum systems can only reach coupling strengths
that are within the TC model \cite {Fink:2009hi,Lucero2012,Mlynek2012},
it would be of significant interest to see if a non-equilibrium QPT can be observed
within a driven TC model \cite{milburn,zou}. In comparison with the Dicke model under a drive, no approximation
is necessary to transfer the driven TC model from the laboratory frame to the rotating frame.
Within the rotating frame, a QPT is indeed predicated \cite{zou} at a critical coupling strength below the Dicke critical
coupling strength. This driven TC QPT critical coupling is comparable to the geometric mean of the spin and field detunings
from the drive frequency (here and below referred to as the TC critical coupling, in comparison with the Dicke critical coupling).

Here we show experimental evidence confirming the existence of such a non-equilibrium QPT in a driven TC circuit,
with four superconducting phase qubits each coupled, at a fixed strength
smaller than the Dicke critical coupling by 200 times, to a superconducting coplanar waveguide resonator.
We witness the non-equilibrium QPT through a dynamical measurement, by recording the time evolution of the
four-spin joint occupation probabilities while the TC critical coupling strength is swept over time.
In the experiment we demonstrate the high-level of control possible in our system
by off-resonantly driving the common resonator mode and subsequently fine-tuning the qubit frequency to cross the TC QPT critical point,
with the results measured at different microwave drive strengths and durations
in good agreement with theory.\\

\noindent\textbf{Results}

\noindent \textbf{The system and the Hamiltonian.} Our driven TC circuit is built in
a circuit-QED configuration~\cite{wallraff2004}, which realizes the on-chip analogue of cavity-QED.
Inheriting the high scalability and controllability from microwave integrated circuits \cite{max,sun}
and benefiting from the significant coherence improvement of
superconducting qubits over the past decade \cite{Barends,rige},
circuit-QED systems based on superconducting qubits and
resonators~\cite{clarke,YouNori} are suitable for building
large-scale quantum simulators~\cite{houck,nori-si-1,nori-si-2,chen} to study
fundamental many-body problems.

Figures~1a and b present the circuit layout, which consists of four superconducting
phase qubits coupled to a common coplanar waveguide resonator \cite {Lucero2012}. The resonator frequency is fixed at $\omega_{\rm r}/2\pi\cong$ 6.2 GHz,
around which the resonance frequency of each qubit ($\omega^{k}_{\rm q}$ for $k$ = 1, 2, 3 or 4) can be individually adjusted.
The resonator's energy decay rate is $\kappa_1 \cong$ 0.4 MHz and its pure dephasing rate $\kappa_2$ is negligible.
Because the energy decay rate and the pure dephasing rate for the qubits slightly vary as functions of qubit frequency,
we sample their values in a frequency range from 6 GHz to 6.15 GHz and take the average in numerical simulation.
The qubits' energy decay rates are, on average, $\Gamma_1 \cong$ 2.0 MHz and their pure dephasing rates are, on average, $\Gamma_2 \cong$ 4.0 MHz.
Couplings between each qubit and the resonator are fixed by designing the coupling capacitors to be nearly identical
and therefore we consider a homogenous coupling strength $\lambda/2\pi =30$ MHz in the following treatment
(Supplementary Note 5 and Fig. 4 for detailed sample parameters).

\begin{figure*}[t]
\centering {\includegraphics[width=12.4 cm, clip=True]{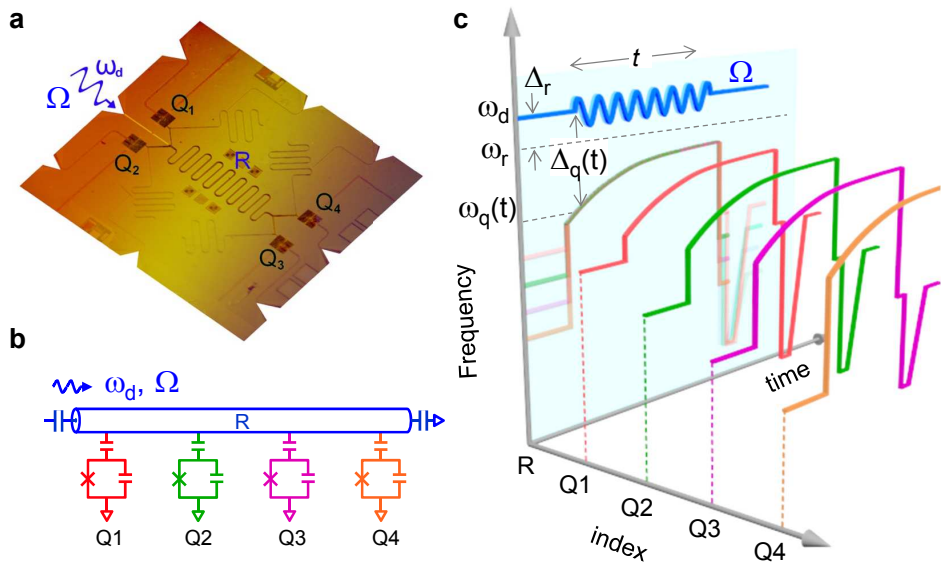}}
\caption{\textbf{Diagrams of device and measurement sequence.}
(\textbf{a}) A false-color device image highlighting the
circuit elements such as the qubits (dark squares) and the half-wavelength coplanar waveguide resonator (the sinusoidal line in the middle).
Four superconducting qubits $Q_k$ ($k$ = 1, 2, 3, and 4) are individually
coupled to $R$. The microwave drive to the resonator is
applied through the transmission line between $Q_1$ and $Q_2$ as indicated.
(\textbf{b}) Simplified circuit schematic.
(\textbf{c}) Illustration of the pulse sequence, where the x-axis indexes the qubits and the resonator,
the y-axis represents the sequence time and the z-axis represents the
frequency (Supplementary Note 5 for designing the sequence). The four qubits, originally sitting at their idling
frequencies, are simultaneously tuned to the same frequency $\omega_{\rm q}(t_0)$
such that $\lambda/\lambda_{\rm c} = 0.5$ (at this point all qubits and the resonator are individually in their own ground state), following which
$\omega_{\rm q}(t)$ is swept for a time $t$ up to $\tau$, such that $\lambda/\lambda_{\rm c}$ increases uniformly from 0.5 to 2.5 over the full period of $\tau$ (see the asymptotic curves and their shades).
During the ramping, a microwave drive (the blue sinusoidal line) to the resonator $R$ with a fixed frequency $\omega_{\rm d}$
and a fixed drive strength $\Omega$ is always on (Methods for determining $\Omega$).
We record the four-qubit occupation probabilities as functions of the sweep time $t$, by simultaneously tuning all four qubits
to their measurement points at lower frequencies for joint qubit-state readout after sweeping $\omega_{\rm q}(t)$ (see the sharp trapezoids and their shades): in each sequence we record each qubit's state by `0' or `1' in a single-shot manner,
and repeating the same sequences many times ($\sim\,10^3$ to $10^4$),
we count the sixteen probabilities $P_{0000}$, $P_{0001}$, $P_{0010}$, $\cdots$, and $P_{1111}$,
where `0' and `1' denote, respectively, the ground and excited states of each qubit.
These probabilities are used to calculate the collective spin operator
$\langle J_{z}\rangle$ (Methods). }
\end{figure*}

Applying an external microwave tone at $\omega_{\rm d}$,
we may generally describe the Hamiltonian of the system in a rotating frame as,
\begin{eqnarray}
H_{0}=\frac {1}{2}\sum_{k=1}^{N} \Delta^{k}_{\rm q}\,\sigma^{k}_{z}+\Delta
_{\rm r}\,a^{\dagger }a+ \sum_{k=1}^{N} \frac{\lambda}{\sqrt{N}}
\,(a\sigma^{k}_{+}+ a^{\dagger}\sigma^{k}_{-}) \notag \\
+\Omega\,(a+a^{\dagger})+ \sum_{k=1}^{N} \frac{\Omega'_{k}}{\sqrt{N}}
\,(\sigma^{k}_{+}+ \sigma^{k}_{-}),  \label{1}
\end{eqnarray}
where $N=4$, $\Delta^{k}_{\rm q}=\omega^{k}_{\rm q}-\omega_{\rm d}$ ($\Delta_{\rm r}=\omega_{\rm r}-\omega_{\rm d}$) is the
detuning of the qubit (resonator) resonance from the drive frequency, $a$ ($a^{\dagger}$) is the
lowering (raising) operator of a single mode of the resonator, and $\sigma^{k}_{\pm},
\sigma^{k}_{z}$ are the $k^{\rm th}$-spin Pauli operators. $\Omega$ and $\Omega'_{k}$ are, respectively,
the driving strengths to the resonator and to the $k^{\rm th}$ qubit. To understand the non-equilibrium QPT as
derived from equation (1), we assume four identical spins for simplicity, and we simultaneously steer all four qubits on
the same frequency trajectory $\omega_{\rm q}(t)$ as the system evolves, yielding $\Delta^{k}_{\rm q}=\Delta_{\rm q}(t)$.
This assumption applies to our experiment (Fig. 8) and
reduces the complexity due to parametric inhomogeneity \cite{Lopez:2007do, Ian2012}.
Under another homogeneous approximation, i.e., $\Omega'_{k}=\Omega'$, the last two terms in equation (1),
regarding the drivings on the resonator and the qubits, are unitarily equivalent \cite{zou}.
Since our microwave tone in the present experiment is designed to drive the resonator, the effect due to driving
the qubits via the unwanted but small microwave crosstalk can be absorbed into that of driving the resonator.
As a result, equation (1) is simplified by neglecting the small terms involving $\Omega'_{k}$ in the following treatment.

The original undriven TC model possesses, in theory, a critical
coupling at $\lambda_{\rm c}^{\rm o}=\sqrt{\omega_{\rm q}\omega_{\rm r}}$, which is impossible to reach in our circuit.
In a rotating-frame variant of the driven TC Hamiltonian shown in equation (1), the qubit (resonator)
resonance frequency $\omega_{\rm q}$ ($\omega_{\rm r}$) is replaced by
the associated detuning $\Delta_{\rm q}$ ($\Delta_{\rm r}$), yielding
a QPT whose critical point now scales with the geometric mean of the spin and field detunings, i.e., $\lambda_{\rm c}=\sqrt{\Delta_{\rm q}\Delta_{\rm r}}$.
For such an off-resonantly driven TC system, the QPT can be traversed by engineering the homogenous coupling strength $\lambda$ to pass through $\lambda_{\rm c}$.

Since $\lambda$ is fixed in our case, to experimentally observe the QPT, we sweep $\lambda_{\rm c}$
through $\lambda$, i.e., we sweep $\Delta_{\rm q}$ such that $\lambda/\lambda_{\rm c}$
increases linearly with time from 0.5 to 2.5 over a duration of $\tau$ (see Fig. 2b inset).
The detailed pulse sequence for performing the experiment is illustrated in Fig. 1c:
starting with all qubits and the resonator in their own ground states, we turn
on the microwave drive at a fixed resonator-drive detuning $\Delta_{\rm r}$ and then
immediately tune all four qubits to the same frequency such that $\lambda/\lambda_{\rm c} = 0.5$. Following this
we sweep the qubit frequency on an asymptotic trajectory (achieving a constant ramping rate for $\lambda/\lambda_{\rm c}$) for a time duration $\tau$.
Dynamics of the system during the ramping of $\lambda/\lambda_{\rm c}$ are measured by recording
the four-qubit joint occupation probabilities as functions of the sweep time.
Evidence of the QPT can be witnessed in the change of the inferred mean values
of the collective spin operator, i.e., $J_{z}=\sum_{k=1}^{4}\sigma^{k}_{z}/2$, as $\lambda/\lambda_{\rm c}$ increases above 1.

We note that our qubit is not an exact spin-1/2 system due to its weak anharmonicity, i.e., there exists
a next higher energy state. The pulse sequence, shown in Fig. 1c, is designed
to avoid significant state population leakage to the qubit's next higher energy state.
When probing the four-qubit dynamics we specifically parametrise some relevant Hamiltonian parameters such as
the drive strength $\Omega/2\pi$, the resonator-drive detuning $\Delta_{\rm r}/2\pi$
and the total sweep duration $\tau$ under experimental constraints (Supplementary Note 5 for detailed discussions).\\

\noindent\textbf{The quantum phase transition in the driven Tavis-Cummings model.}
Before presenting our experimental results, we first describe the ideal QPT as predicted by theory \cite{zou},
and relate it to our experimental reality. In particular we try to clarify the quantum critical behaviour in the context of a few qubits
coupled to a common resonator mode and discuss the connection between the few-qubit case and the case in the thermodynamic limit;
we also try to clarify how the dynamical measurement via a swept $\omega_{\rm q}(t)$ (equivalent to uniformly varying
$\lambda/\lambda_{\rm c}$ over time), correlates with a signature of the QPT.
According to \cite{zou}, the QPT is present when the system switches from a normal phase to a superradiant phase in the rotating frame.
In this generic ground state QPT, around the QPT's critical point ($\lambda/\lambda_{\rm c}=1$)
we may observe a sharp cusp in the scaled moments $\langle J_{x}\rangle/(N/2)\sim|\lambda/\lambda_{\rm c}-1|^{\gamma_{x}}$ (with
$\gamma_{x}=1/2$) and $\langle J_{z}\rangle/(N/2)\sim|\lambda/\lambda_{\rm c}-1|^{\gamma_{z}}$ (with $\gamma_{z}=1$)
for $\lambda/\lambda_{\rm c}\ge 1$,
and also in the mean number of photons with $\langle a^{\dagger}a \rangle /(N)\sim |\lambda/\lambda_{\rm c}-1|^{\gamma_{a}}$ (with $\gamma_{a}=1$) for $\lambda/\lambda_{\rm c}\ge 1$. The critical exponents $\gamma_{x,z,a}$ represent the critical scaling behaviour observable in the thermodynamical limit (Supplementary Note 1).
In contrast to the cusp-like behaviour in the thermodynamic limit, the QPT in the few-qubit case yields
$\langle J_{z}\rangle/(N/2)$ curves that rise in a smooth but abrupt manner (Fig. 2a). Nevertheless, the
critical point for the few-qubit case can still be visually identified proximal to $\lambda/\lambda_{\rm c} = 1$.
Due to the dissipative nature of our system and the hardware limitation we focus our observation on
$\langle J_{z}\rangle/(N/2)$, whose behaviour around $\lambda/\lambda_{\rm c} = 1$
can be a sufficient evidence of the QPT (Discussion and Methods).

\begin{figure*}[tbph]
\centering {\includegraphics[width=10.4 cm]{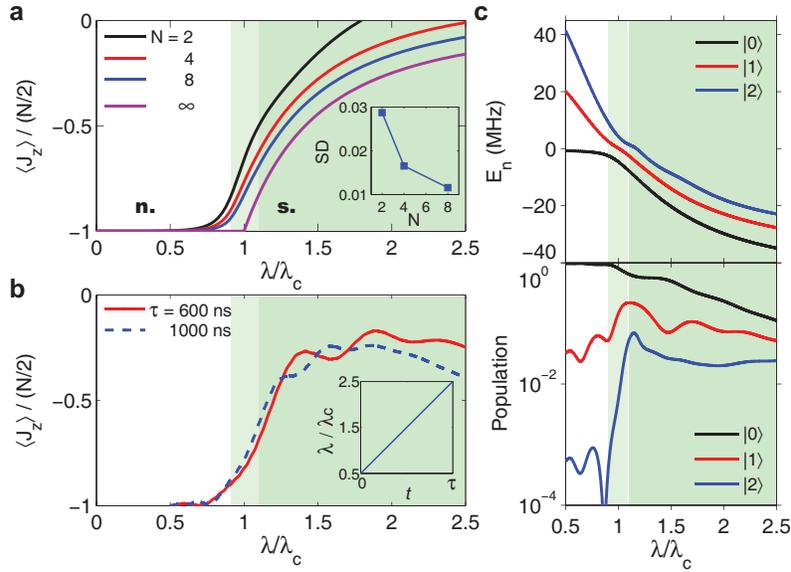}}
\caption{\textbf{$\langle J_{z}\rangle$'s signature behaviours across the critical point
in the ground state quantum phase transition and in the experimental dynamics}, calculated with $\Delta_{\rm r}/2\pi= 30$~MHz and $\Omega/2\pi=4$~MHz for example.
The quantum critical region, illustrated by the light-green background
in all panels, happens around $\lambda/\lambda_{\rm c}=1$ between the normal ({\bf{n.}}, the white region) and superradiant ({\bf s.}, the green region) phases.
(\textbf{a}) Numerical calculations of $\langle J_{z}\rangle/(N/2)$ by solving equation (1) for the ground state
at different number of qubits $N$ as indicated.
The cusp-like behaviour at the critical point $\lambda/\lambda_{\rm c}=1$ occurs only in the thermodynamical limit, and
the finite qubit cases ($N$ = 2, 4 and 8) display the drastic rise after traversing the critical point. Inset illustrates
the maximal standard deviations (SD) of $\langle J_{x}\rangle/(N/2)$ as calculated in ({\bf a}) due to
random noise (or inhomogeneity) for different number $N$ of qubits involved. For illustrative purpose here
we only consider the frequency uncertainty in each qubit $\delta[\omega_{\rm q}^{k}/2\pi]=\pm 1$ MHz, relevant to our experimental setup. It is seen that uncertainties do not give large errors
and increasing the number of qubits yields better suppression of the random noise.
(\textbf{b}) Numerical calculations of $\langle J_{z}\rangle/(N/2)$ as function of $\lambda/\lambda_{\rm c}$
following the experimental pulse sequence in Fig. 1c at different durations as indicated. Sample decoherence is included in calculations.
It is seen that $\langle J_{z}\rangle/(N/2)$ curves rise around the same point as that in (\textbf{a}).
Inset illustrates $\lambda/\lambda_{\rm c}$ as a function of the ramping time $t$ during the pulse sequence.
(\textbf{c}) Numerically calculated energies of the lowest three energy eigenstates (top) and
population distribution (in logarithmic scale) among these three states (bottom) of the four-qubit Hamiltonian system described in equation (1)
as functions of $\lambda/\lambda_{\rm c}$ under decoherence, with $\tau=600$~ns. Higher energy states are omitted for clarity.
Starting with all qubits and the resonator in their own ground states at $\lambda/\lambda_{\rm c} = 0.5$
(at this point the system's ground state $\left|0\right\rangle$ is at $E_0 \approx 0$ and takes the largest population as shown by the black line),
$E_n$ of the lowest few states significantly drop below 0 and the population distribution quickly evolves as $\lambda/\lambda_{\rm c}$ increases above 1 (in the light-green region),
indicating a structural change of the eigenstates of the system crossing this critical point.}
\end{figure*}

The QPT as evidenced in Fig. 2a, by the rise of $\langle J_{z}\rangle/(N/2)$, is a generic ground state QPT in the rotating frame \cite{zou}.
Starting from the normal ground state at $\lambda/\lambda_{\rm c} \ll 1$, to reach the superradiant ground state at $\lambda/\lambda_{\rm c} > 1$
we have to ramp up $\lambda/\lambda_{\rm c}$ very slowly, in accordance with the adiabatic condition. For an open quantum system the adiabatic
condition can be difficult to satisfy since dissipation plays a decisive role given long enough evolution times.
As such, we examine the QPT in a non-adiabatic manner: we ramp up $\lambda/\lambda_{\rm c}$ quickly and linearly over durations that range
from a few hundred to a thousand nanoseconds (comparable with the qubit energy relaxation time 1/$\Gamma_1$), in order to minimize the
impact of dissipation on the dynamics. During the process we constantly monitor the four-qubit occupation probabilities, from which
we calculate $\langle J_{z}\rangle/(N/2)$ to study its behaviour over time. As a comparison,
we numerically model the time evolution of $\langle J_{z}\rangle/(N/2)$ under open system
dynamics as described by equation (1) based on a master equation approach (Supplementary Note 3).
As our numerical simulation suggests, excited states of the system
can be populated during the evolution, and the population distribution among different eigenstates
tends to stabilize after $\lambda/\lambda_{\rm c}$ increases above 1.5 (Fig. 2c).
In particular, in the non-adiabatic process and under decoherence, $\langle J_{z}\rangle/(N/2)$
still rises up around $\lambda/\lambda_{\rm c}=1$, in a style (Fig. 2b) very similar to that in Fig. 2a.
Therefore, the onset where $\langle J_{z}\rangle/(N/2)$ rises up abruptly from -1 should correlate well with
the critical point of the generic ground state QPT, which in itself reflects a situation where a qualitative
change occurs in the properties of the system's eigenstates as
a function of the Hamiltonian parameter in equation (1) (here $\lambda_{\rm c} = \sqrt{\Delta_{\rm q}\Delta_{\rm r}}$).
Our experiment, though involving the system's higher energy states, should still provide strong
evidence for the QPT via the observed abrupt change of $\langle J_{z}\rangle/(N/2)$ as $\lambda/\lambda_{\rm c}$
is tuned through unity.\\

\noindent\textbf{Experimental observation of the quantum phase transition.}
Following the experimental sequence outlined in Fig. 1c, in Fig. 3a we show
the typical dynamics measured at $\Delta_{\rm r}/2\pi = -30$~MHz, $\Omega/2\pi = 4$~MHz and $\tau = 600$ ns, with
the sixteen four-qubit joint occupation probabilities ($P_{0000},\,P_{0001},\,P_{0010},\,\cdots$) evolving with the sweep time $t$.
The choice of a negative $\Delta_{\rm r}$ is to minimize the state leakage caused by the microwave drive, which
should not affect the dynamics and the QPT physics as calculated using a positive $\Delta_{\rm r}$ in Fig. 2.
The sixteen probabilities can be grouped according to their excitation quanta, and
the very close dynamics of the probabilities in the same group suggest that
four qubits behave similarly, validating the identical spin assumption in the QPT theory.
$\langle J_{z}\rangle/(N/2)$ can be calculated using these sixteen probabilities, as processed in Fig. 3b (points with error bars).
By mapping the x-axis in time to $\lambda/\lambda_{\rm c}$, we display the $\langle J_{z}\rangle/(N/2)$ versus $\lambda/\lambda_{\rm c}$ curves,
with values of $\Omega$ as listed and $\Delta_{\rm r}/2\pi = -30$~MHz, for the cases of $\tau = 600$ and 1000 ns in Figs. 3c and d, respectively
(more experimental data for $\Delta_{\rm r}/2\pi = -20$~MHz can be found in Fig. 5).
Comparing with those shown in Figs. 2a and b, it is seen that the experimental results have unambiguously caught the main feature of the off-resonantly driven
QPT, i.e., a signature rise of the scaled moment $\langle J_{z}\rangle$ as $\lambda/\lambda_{\rm c}$ increases above 1, the critical point.
We note that the spectral line-widths for the qubits and the resonator are defined by their energy relaxation and dephasing rates ($\Gamma_1$, $\Gamma_2$, $\kappa_1$, and $\kappa_2$),
all less than values of $|\Delta_{\rm q}|$ (e.g., $\approx 2\pi\times 13$~MHz at $\lambda/\lambda_{\rm c} = 1.5$ where $\langle J_{z}\rangle$ rises to a high level. Note that $\Delta_{\rm q}<0$) and $|\Delta_{\rm r}|$ ($=2\pi\times 30$~MHz) used in measurements for data in Fig. 3.
We also verify that state leakage to the next higher energy state of the qubits is reasonably small
during these measurements (Fig. 6).
As such, the rise around $\lambda/\lambda_{\rm c} = 1$ is compatible with
the critical point quoted in the context of the non-equilibrium QPT, which reflects a structural change of the system's eigenstates.

\begin{figure*}[t]
\centering {\includegraphics[width=12.4 cm]{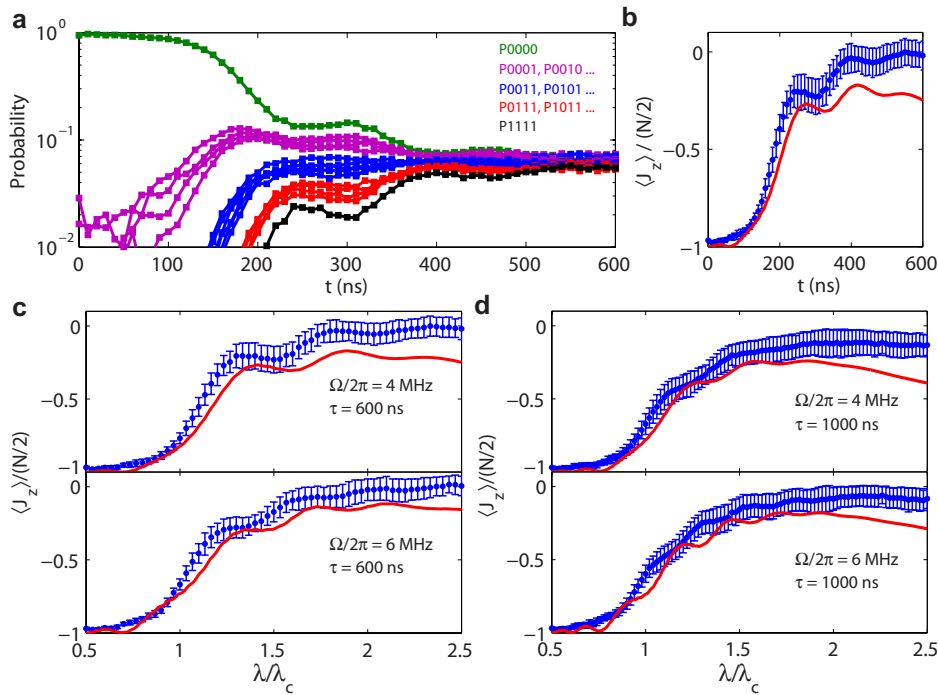}}
\caption{\textbf{The four-qubit quantum phase transition experimental results in comparison with numerical simulation}.
(\textbf{a}) Four-qubit joint occupation probabilities (in logarithmic scale) as functions of the sweep time $t$
for $\Delta_{\rm r}/2\pi = -30$~MHz, $\Omega/2\pi = 4$~MHz and $\tau = 600$~ns
(error bars, on the order of 1\%, are not shown for clarity). $\lambda_c/2\pi$
varies from 60 MHz at $t$ = 0 ns to 12 MHz at $t$ = 600 ns. The sixteen probabilities
are grouped by their corresponding excitation quanta as marked by different colors:
green for no excitation ($P_{0000}$), purple for one quantum excitation  ($P_{0001}$, $P_{0010}$, $P_{0100}$, $P_{1000}$),
blue for two quanta excitation ($P_{0011}$, $P_{0101}$, $\cdots$, $P_{1100}$),
red for three ($P_{0111}$, $P_{1101}$, $P_{1011}$, $P_{1110}$) and black for four ($P_{1111}$).
The critical point is approached when the two-quanta-excitation curves start to gain finite probability values.
Curves in the same group behave similarly, validating the identical spin assumption in the QPT theory.
(\textbf{b}) $\langle J_{z}\rangle/(N/2)$ dynamics calculated from data in (\textbf{a}) (points with error bars).
Line is a numerical simulation.
Error bars are standard deviations of repetitive measurements, during each measurement we add a random bias sequence to
each qubit to simulate the frequency uncertainties of $\pm1$~MHz (the uncertainty level of our calibration of the qubit frequency).
Experimentally measured error bars agree with numerical calculations considering all known uncertainties in our experiments,
with the majority of the errors coming from the frequency uncertainties in biasing the qubits, which accumulate
over the sweep time, and the readout uncertainties of the occupation probability.
(\textbf{c, d}) $\langle J_{z}\rangle/(N/2)$ as functions of $\lambda/\lambda_{\rm c}$ showing the existence of QPT (points with error bars).
Lines are numerical simulations including decoherence. Error bars are obtained similarly to those in ($\textbf{b}$).
The choice of a negative $\Delta_{\rm r}$ is to experimentally minimize the state leakage, which
should not affect the dynamics and the QPT physics as calculated using a positive $\Delta_{\rm r}$ in Fig. 2.
Experimental signal of $\langle J_{z}\rangle$ is slightly larger than theory prediction
due to the slight state leakage which is miscounted as $\langle J_{z}\rangle$'s signal in the measurements (Figs. 6 and 7). }
\end{figure*}

Different from the ideal QPT case in the isolated system, the interplay between the external drive
and decoherence irreversibly evolves the system into a non-equilibrium quasi-steady state,
where the term quasi refers to the fact that $\langle J_{z}\rangle/(N/2)$ tends to approximately level off at longer sweep times $\tau$.
Using typical coherence parameters of our device we also simulate the experimental conditions for the experimental data shown in Figs. 3b-d (lines).
The numerical results show that the system reaches the quasi-steady state approximately after $\lambda/\lambda_{\rm c}>$1.5, and the situation slightly varies with $\Delta_{\rm r}$.
Nevertheless, our experimental data are in good agreement with numerical simulations taking into account decoherence with no fitted parameters.

Moreover, a faithful simulation requires a clear understanding of the operational imperfections,
particularly when the underlying problem is otherwise intractable.
As discussed in the Supplementary Note 5, we specifically design the pulse sequence to
minimize the dominant experimental imperfections in equation (1), including using appropriate negative detunings of \{$\Delta_{\rm r}$, $\Delta_{\rm q}$\}
and ramping rates of $\lambda/\lambda_{\rm c}$ (or equivalently the sweep durations $\tau$).
Nevertheless, there are other experimental subtleties that we cannot avoid, e.g.,
slight state leakage (miscounted as $\langle J_{z}\rangle$'s signal in the measurements).
By taking into account of the state leakage, we find better agreement between the experimental data
and theory (details in the Supplementary Note 5).\\

\noindent\textbf{Discussion}\\
To further understand quantum critical behaviour beyond the observed QPT, we have to return to the noiseless model.
We first refer to an undriven TC model in comparison with the driven TC model (equation (1)), in the latter of which
the ground state QPT  is related to a breaking of the parity symmetry \cite{zou}.
The undriven TC Hamiltonian $H_{\rm tc}= \omega_{\rm q}\,J_{z} + \omega_{\rm r}\,a^{\dagger}a + \frac{\lambda}{2} \,(a J_{+} + a^{\dagger} J_{-})$
commutes with a parity operator $P=e^{i\pi L}$, where $L= J_{z} +a^{\dagger}a + N/2$ represents the total number of
excitations of the collective system. Parity conservation ensures
that $\langle J_{x}\rangle$ remains zero and $\langle J_{z}\rangle$
increases in a staircase fashion as the spin-field coupling
$\lambda$ increases. With increasing $\lambda$, it is shown \cite{zou} that level crossings
occur, e.g., between the ground state and the excited states, after crossing the
critical point (i.e., $\lambda>\sqrt{\omega_{\rm q}\omega_{\rm r}}$ purely for theoretical discussion only.
Note that by definition $\lambda \ll \sqrt{\omega_{\rm q}\omega_{\rm r}}$ is required in the TC model), and
this results in discrete parity changes in the state of the system.
However, in the driven TC model, with critical point changed from $\sqrt{\omega_{\rm q}\omega_{\rm r}}$ to
$\sqrt{\Delta_{\rm q}\Delta_{\rm r}}$, the QPT really happens, but parity is no longer conserved since $[H_{0}, P]\neq$ 0.
The broken parity can be associated with avoided level crossings in
the eigenspectra, and the excitation number $L$ is no longer
conserved. This results in a smooth rise, i.e., a 'rounded'
staircase for $\langle J_{z}\rangle$, after crossing the critical
point, if the microwave drive is weak enough (note that in the present experiment
the weak drive limit is not reached since equation (1) assumes four identical spins, which requires that
the drive strength $\Omega$ has to cover, at least, the frequency uncertainties
while simultaneously tuning all four qubits to follow the same frequency trajectory $\omega_{\rm q}(t)$).

The QPT under consideration also behaves differently from the Dicke QPT \cite{Baumann:2010js} as regards parity.
It is a generic Dicke model, rather than a driven Dicke model, achieved in  \cite{Baumann:2010js}, for which parity
is conserved in the normal phase. In contrast, no parity is conserved in our driven TC system
in either the normal or the superradiant phase. Parity breaking is
responsible for some scaling behaviour and the parity symmetry is relevant
to various types of quantum correlations \cite{lambert, buzek}.

The driving in our case not only breaks the parity symmetry of the original TC system, but also helps circumventing
the 'no-go' theorem due to the restriction
from the Thomas-Reiche-Kuhn sum rule \cite{nogo1,nogo2,nogo3}. As discussed in Supplementary Note 2,
the small $A^{2}$ term (with $A^{2}=\kappa(a+a^\dagger)^2$), which is neglected in the above treatment but whose appearance
might forbid the QPT, turns to be a harmless shift in $\Delta_{\rm r}$ in the rotating frame.
Therefore, the introduction of the driving profoundly alters the resulting physics, enabling
the observation of a non-equilibrium QPT.

Although our data are fully compatible with the non-equilibrium QPT picture as predicted
by the off-resonant driven TC theory, it is worth noting that currently we cannot exclude the possibility
of a semi-classical interpretation of our experiment.
In contrast to our experimental condition that involves only finite quantum elements and finite excitation levels in each element,
semi-classical treatments employ continuous variables which would work better in the thermodynamic limit.
Unfortunately, the relevant semi-classical treatments that we are aware of only deal with specific conditions and
are not suitable for interpreting our off-resonant driven TC experiment (Supplementary Note 4).
Therefore, whether a semi-classical alternative is possible to explain the data remains an open question.
Along this route, the tomography measurement of the QPT dynamics, though technically challenging,
could allow the exploration of any possible quantum correlations encoded in the QPT, which would be useful to
answer the open question as regards a semi-classical alternative in future experiments.

In addition, following on from this work we expect demonstrations of a staircase behaviour in $\langle J_{z}\rangle$
and a cusp-like behaviour in $\langle J_{x}\rangle$ around the critical point, both hallmarks of the generic ground state QPT, in future experiments
using larger numbers of closely identical qubits with improved coherence and more sophisticated control. By further
suppressing decoherence we may enable the demonstration of the ground state QPT in a configuration similar to the present device
strictly following the proposed implementation in \cite{zou}. With recent progress in
superconducting quantum information technology and the promising outlook to develop
intermediate-scale complex quantum circuits, we believe that further
exploration of many-body physics in a non-equilibrium condition by building a solid-state quantum simulator
with only weak spin-field couplings can be expected in the near
future. This will help improve our understanding of the interplay between non-equilibrium and
quantum correlations as well as the role of parity symmetry in many-body systems.\\

\noindent\textbf{Methods}

\footnotesize{\textbf{Tuning $\Omega$.} The microwave drive strength
$\Omega$ in equation (1) depends on both the microwave drive
amplitude $A$ and the coupling capacitance for feeding energy into
the resonator, with the latter being set once the device and the
measurement setup are fixed. $A$ is what we usually quote using the room-temperature electronics.
More importantly, the $\Omega$-$A$ relation could weakly depend on the off-resonance magnitude of $\Delta_{\rm r}$,
due to the frequency dependence of the transmission coefficient of
the microwave cables at cryogenic temperatures, and the interference by various box modes and spurious two-level defect modes.
To find out the exact $\Omega$ used in our experiment, we start with determining the on-resonance $\Omega$
by calibrating the relation between $\Omega$ and $A$.
We resonantly drive the resonator using a single-tone microwave
pulse with an amplitude $A$ for a period of $t$ (typically 50
ns), after which we bring a qubit, originally in its ground state,
to resonantly interact with the resonator for detecting the
resonator state~\cite{max}. The microwave drive generates a
coherent state in the resonator and the subsequent qubit-resonator
interaction results in multi-tone vacuum Rabi oscillations
whose frequencies depend on the resonator populations. We record
the time evolution of the qubit probabilities in the excited state
for the first 300 ns, from which the energy-level population
probabilities of the resonator ($P_n$ for $n$ = 0, 1, 2, $\cdots$) are
inferred. The resonator energy-level populations satisfy the Poisson
distribution and we quote the displacement $\alpha$, which is the
square root of the average photon number in the resonator, by
$\alpha = \left(\sum_{n}{n P_n}\right)^{1/2}$. For a
fixed experimental setup and a fixed $t$, the ratio $\gamma =
\left|\alpha\right|/A$ is a constant, and can be experimentally
determined by sampling a group of $A$ and $\alpha$ values. The drive
strength is thus $\Omega = \gamma A/t$.

To calibrate the off-resonance $\Omega$, we carry out the 2-qubit experiment
with a similar setup as discussed in Fig. 1, using the on-resonance
$\Omega$ value $(=2\pi\times 4$ MHz) as an initial trial.
The measured results are then compared with numerical simulation,
which verifies that $\Omega$ values calibrated on resonance
are also applicable at small detuning values of $\Delta_{\rm r}$ used in the
four-qubit experiment (Fig. 9 for more detail).

\textbf{Qubit readout and the correction.} The qubit readout is done
using an integrated superconducting quantum interference device
(SQUID), which can tell the flux difference between the qubit's
ground and excited states. The readout details can be found in
Ref.~\cite{linpeng}. We simultaneously read out all the states of the
four qubits (one SQUID for each qubit), therefore obtaining the
sixteen probabilities {$P_{0000}$,
$P_{0001}$, $P_{0010}$, ..., $P_{1111}$}. These values are corrected before
further processing. The readout fidelities for $|g\rangle$ ($F_{k,g}$)
and $|e\rangle$ ($F_{k,e}$) for qubit $Q_k$ are obtained using the single-qubit measurement.
The correction matrix for $Q_k$ is the inverse of
\begin{equation}
F_{k} = \left[ \begin{matrix} F_{k,g} & 1-F_{k,e} \\ 1-F_{k,g} & F_{k,e} \end{matrix} \right].
\end{equation}
We correct all sixteen values using the inverse of the
tensor-product matrix $F_{1} \otimes F_{2} \otimes F_{3} \otimes F_{4}$.
The correction matrix may be slightly off due to the small flux crosstalk when simultaneously reading
out all four qubits, which is a possible reason for
that the experimental $\langle J_z \rangle /(N/2)$ value does not start from
-1.0 at $\lambda/\lambda_{\rm c} = 0.5$ in Fig.~3.

\textbf{Expectation values of the spin operator $J_{z}$.} After
correcting the sixteen qubit-state probabilities, we calculate the scaled
$\langle J_{z}\rangle$ using
\begin{equation}
\langle J_{z}\rangle/(N/2)=\sum_{i_1,i_2,i_3,i_4 = 0,1}\left(\frac{i_1+i_2+i_3+i_4}{2}-1\right)P_{i_1,i_2,i_3,i_4},
\end{equation}
where $i_k$ = 0, 1 represents the ground and excited states of qubit
$Q_k$, respectively, and the summation runs over all four-qubit
eigenstates corresponding to the sixteen probabilities.

To calculate $\langle J_{x}\rangle/(N/2)$, the four-qubit state tomography
must be performed, which requires $\pi/2$ rotations on all four qubits (in addition to the microwave tone on the resonator) before
readout. We are unable to measure $\langle J_{x}\rangle/(N/2)$
mainly due to our limited hardware resource. In addition, the involved dynamical
phase when performing the tomography could cause extra complexity in calculating $\langle J_{x}\rangle/(N/2)$.
Since $J_z$ is not affected by the dynamical phase and its rise traversing $\lambda/\lambda_{\rm c} = 1$ can be sufficient proof of the QPT,
we choose to only measure $\langle J_{z}\rangle/(N/2)$ in the experiment.

{\bf Acknowledgments}
We thank J. M. Martinis and A. N. Cleland for providing the device used in the experiment.
This work is supported by National Fundamental Research Program of China (Grant Nos. 2012CB922102 and 2014CB921201), National Natural Science Foundation of China (Grant Nos. 11274352, 11147153, 11274351, 11004226 and 11222437), Zhejiang Provincial Natural Science Foundation of
China (Grant No. LR12A04001) and the Australian Research Council's Centre of Excellence in Engineered
Quantum Systems CE110001013. H.W. acknowledges partial support by National Program for Special Support of Top-Notch Young Professionals.

\end{document}